\documentclass[letterpaper, 10 pt, conference]{ieeeconf}  

\IEEEoverridecommandlockouts                              

\def\snr{\textnormal{SNR}}
\def\im{\, \imath \,}
\def\out{\textnormal{out}}

\usepackage{verbatim}
\usepackage{graphics} 
\usepackage{epsfig} 
\usepackage{times} 
\usepackage{amsmath} 
\usepackage{amssymb}  

\newtheorem{definition}{Definition}[section]
\newtheorem{thm}{Theorem}[section]

\newtheorem{exam}{Example}[section]

\newtheorem{remark}{Remark}[section]

\newcommand{\Z}{{\mathbf Z}}
\newcommand{\R}{{\mathbf R}}
\newcommand{\C}{{\mathbf C}}

\newcommand{\OO}{{\mathcal O}}

\newcommand{\Q}{{\mathbf Q}}

\title{\LARGE \bf
Algebraic Codes and a New Physical Layer Transmission Protocol for Wireless Distributed Storage Systems}

\author{Amaro Barreal$^{1}$, Camilla Hollanti$^{1}$, David Karpuk$^{1}$, and Hsiao-feng (Francis) Lu$^{2}$ 
\thanks{*This work was supported by the Magnus Ehrnrooth Foundation, Finland, through grants to  C. Hollanti and D. Karpuk, and by the Academy of Finland, through a grant to D. Karpuk.}
\thanks{$^{1}$A. Barreal, C. Hollanti, and D. Karpuk are with the Department of Mathematics and Systems Analysis, Aalto University, P.O. Box 11100, FI-00076 AALTO, Espoo, Finland
        {\tt\small firstname.lastname@aalto.fi}}%
\thanks{$^{2}$ Hsiao-feng (Francis) Lu is with the Department of Electrical and Computer Engineering, National Chiao Tung University,
        Taiwan
        {\tt\small francis@mail.nctu.edu.tw}}%
}

\begin{document}

\maketitle
\thispagestyle{empty}
\pagestyle{empty}

\begin{abstract}

In a wireless storage system, having to communicate over a fading channel makes repair transmissions prone to physical layer errors. The first approach to combat fading is to utilize the existing optimal space-time codes. However, it was recently pointed out that such codes are in general too complex to decode when the number of helper nodes is bigger than the number of antennas at the newcomer or data collector. In this paper, a novel protocol for wireless storage transmissions based on algebraic space-time codes  is presented in order to improve the system reliability while enabling feasible decoding. The diversity-multiplexing gain tradeoff (DMT) of the system together with sphere-decodability even with low number of antennas are used as the main design criteria, thus naturally establishing a DMT--complexity tradeoff. It is shown that the proposed protocol outperforms the simple time-division multiple access (TDMA) protocol, while still falling behind the optimal DMT.

\end{abstract}

\section{INTRODUCTION}

Productive and active research in the area of data storage has taken place across several disciplines during the past few years. Recently, attention has been paid to  \emph{distributed  storage systems (DSSs)}, where information is no longer stored in a single device, but rather distributed among several storage nodes in a network, which is potentially wireless. One of the main advantages of storing information in a distributed manner is that the storage system can be made robust against failures by introducing some level of redundancy. This way, even if nodes fail, e.g., go offline, the original file can still be recovered from the surviving storage nodes. In contrast, if a file is stored in a single device and the device breaks, the file is usually lost. Distributed storage also enables the usage of cheap and small disks on small devices instead of having to constantly increase the size of the storage space on an individual device every time the amount of information to be stored increases.  The construction and analysis of a robust distributed storage system is highly nontrivial, and requires a detailed mathematical description. There are several aspects of such a system that need to be considered. 

The first aspect is that of the storage code. To profit from storing data in a distributed manner, a certain amount of redundancy needs to be introduced to the system in order to make it more robust. The most straightforward scenario is replication. This way, if a node is lost, the original information can be recovered by contacting only one of the surviving nodes, provided that all the nodes store a replica. However, this protocol requires to transfer the entire file for the repair of the lost node, and is also not efficient in terms of storage space, as every storage node has to store the whole file. 

More sophisticated protocols have been developed, always giving a tradeoff between the amount of data that needs to be stored in any of the storage nodes, and the amount of data that needs to be retrieved for the repair of a lost node. For more information, see e.g.  \cite{DGWWR07, DRWS10, rashmietal, RR10}. Some examples of real-life distributed storage systems are Apache Cassandra developed at Facebook, and Windows Azure created by Microsoft. 

A second aspect that needs consideration is communication over the existing wireless channels, as it should be possible to store or retrieve a file using a wireless connection \cite{CEET}. The mobility of a user has become crucial in everyday life, and we use wireless channels for data transmission instead of wired ones for increased flexibility. Unfortunately, the transmission of data across wireless channels is risky in terms of transmission errors, as waves traveling through the channel typically suffer from several effects imposed by the nature.

From the mathematical point of view, this process of transmission can be significantly improved by using  suitable coding techniques, a prominent one being space-time (ST) coding. Space-time codes are used when transmitting information over a wireless multiple-input multiple-output (MIMO) channel between terminals equipped with multiple antennas, and they provide a certain amount of redundancy by using several time slots for the transmission of the same information. A space-time code is a finite subset of $n_t\times T$ complex matrices, where $n_t$ is the number of transmit antennas and $T$ the number of channel uses. The main design criteria for ST construction are 1) the rank criterion, 2) the determinant criterion, and 3) the diversity-multiplexing gain tradeoff (DMT) \cite{TSC, ZT, ZT_MAC}. For a good survey on algebraic space-time codes we refer to \cite{OV,oggierviterbo2}. As a recent example, Apple's new iPad Air, released November 2013, is equipped with two antennas instead of one and profits from the advantages of MIMO coding techniques, increasing the speed of data transmission as well as the reliability of its wireless communications.

\subsection{Contributions and related work}

In most of the storage related research the focus is on the (logical) network layer, while the physical layer is usually ignored. Nonetheless, some interesting works considering the physical layer do exist. In \cite{gong}, a so-called partial downloading scheme is proposed that allows for data reconstruction with limited bandwidth by downloading only parts of the contents of the helper nodes. In \cite{Ning}, the use of a forward error correction code (\emph{e.g.}, LDPC code) is proposed in order to correct bit errors caused by fading. In \cite{rashmi_erasure}, optimal storage codes are constructed for the error and erasure scenario.  

Recently, \emph{space--time storage codes} were introduced in \cite{ICC14} as class of codes that should be able to resist fading of the signals during repair transmissions, while also maintaining the repair property of the underlying storage code. It was pointed out that the obvious way of constructing such codes, namely combining an optimal storage code and an optimal space-time code results in infeasible decoding complexity, when the number of helper nodes is bigger than the number of antennas at the newcomer or data collector (DC). Motivated by this work, we tackle the complexity issue and design a novel yet simple protocol that has feasible decoding complexity for any number of helpers, while the helpers and the newcomer/DC are only  required to have at most two antennas each.

Thus, the present paper deviates from earlier work on DSSs in that it addresses the actual \emph{encoding} of the transmitted data in order to fight the effects caused by fading, continuing along the lines of \cite{ICC14}. In addition to encoding, a sphere decodable transmission protocol for the encoded data will be studied. To make our system as realistic as possible, the protocol requires only up to two antennas at each end while still being sphere-decodable. This is in contrast to the previous optimal space-time codes \cite{kuser} that would be in principle suitable for storage transmissions, but would have exponential complexity when the number of helpers is bigger than the number of receive antennas.   Furthermore, It is shown that the designed system achieves a significantly higher DMT than the TDMA protocol. The observed gap to the optimal DMT is due to the above complexity requirement, which establishes a natural DMT-complexity tradeoff.

\subsection{Distributed storage systems}

In a distributed storage system, the data is stored over $n$ storage nodes by using an $(n,k)$ erasure code. If a maximum distance separable (MDS) code is used to introduce redundancy in the system, then a data collector can reconstruct the whole file by contacting any $k$ out of $n$ nodes. In addition to storing the file, the system has to be repaired by replacing a node with a new one whenever some node fails. This can be done by using, \emph{e.g.}, \emph{regenerating codes} \cite{DGWWR07}. If the \emph{newcomer} node replacing the failed node has to contact  $d$ \emph{helper} nodes in order to restore the contents of the lost node, we call the code an $(n,k,d)$ code, and $d$ is referred to as the \emph{repair degree}. Recent work \cite{DGWWR07,me} considers tradeoffs between the storage capacity, secrecy capacity, and repair bandwidth. Explicit storage code constructions achieving some of the tradeoffs can be found in \cite{RR10, DRWS10, rashmietal}, among many others. Regenerating codes by definition achieve the storage capacity--repair bandwidth tradeoff.

\section{Storage Communication over Parallel Fading Channels}
In this section, we review from \cite{ICC14} the basic idea of space-time coded storage transmissions. If a node fails in the system or a data collector (DC) wants to reconstruct the stored file, then the newcomer/DC\footnote{We denote both by $u_{new}$ for simplicity.} node $u_{new}$ has to contact $K\in\{d,k\}$ nodes for repair/reconstruction.  We denote by
$
u_{i_1},\ldots,u_{i_K}
$
 the nodes contacted by $u_{new}$.  Each $u_{i_j}$ would like to send its contents $\bar{x}_{i_j}$ to $u_{new}$ over a Rayleigh fading channel.  In order to match the alphabets used for the stored information on one hand, and for the wireless transmission on the other hand, a bijective lift function is introduced to the system. Formally,
\[
L:\mathcal{X}\rightarrow \mathcal{C}, 
\]
where $\mathcal{X}$ is the set of possible encoded\footnote{Encoded by an MDS or other erasure code.} file fragments, and $\mathcal{C}$ is a finite constellation of size equal to the size of $\mathcal{X}$. We define $L(\bar{x}_i) = x_i$. To give some intuition to the problem at hand, we start by assuming that the channels used for communication are parallel to each other, e.g., via orthogonal frequency division multiplexing (OFDM) coding,  and that $u_{new}$ has $K$ receive antennas.  The newcomer $u_{new}$  receives the system of equations
\begin{eqnarray*}
y_{i_j} &=& h_{i_j} x_{i_j} + w_{i_j}\,,\quad j=1,\ldots K, 
\end{eqnarray*}
where $h_j$ and $w_j$ are the random complex Gaussian variables describing the channel effect and noise, respectively. The newcomer then decodes these equations to produce ML-estimates
$
\hat{x}_{i_1},\ldots,\hat{x}_{i_K} \in \mathcal{C}
$
of the symbols transmitted over the Rayleigh fading channel with fading components $h_i$.  We then compute
$
\hat{\bar{x}}_{i_j}:= L^{-1}(\hat{x}_{i_j})
$
for all $j=1,\ldots,K$, which are the encoded file fragments actually received by the incoming node $u_{new}$.  To complete the repair, the node $u_{new}$ performs the reconstruction algorithm required by the DSS on these file fragments. This setting can be formalized and generalized as follows:

\begin{definition}[\cite{ICC14}]\label{ST-storage}
A \emph{space-time storage code} consists of the following:
\begin{itemize}
\item[$\bullet$] a DSS system with parameters defined as above, employing an $(n,k)$ MDS code or some other type of erasure code,
\item[$\bullet$] a constellation $\mathcal{C}$ carved from   $\R^{2s}$ or $\C^s$,
\item[$\bullet$] a bijective lift function
\[
L:\mathcal{X}\rightarrow \mathcal{C}, 
\]
where $\mathcal{X}$ is the set of possible encoded file fragments, and
\item[$\bullet$] a space-time transmission protocol using $\mathcal{C}$ as information symbols.
\end{itemize}
\end{definition}

\begin{remark} The storage transmission protocol proposed in this paper makes use of algebraic space-time codes, and hence our 
 constellation $\mathcal{C}$ will be carved from an algebraic lattice  $\Lambda\subseteq\R^{2s}$, and a lift function of the form
\[
L:\{\text{encoded file fragments $\bar{x}_i$}\}\rightarrow \mathcal{C}\subset\Lambda\subset\R^{2s}
\]
will be used. However, the above definition is more general and does not restrict the structure of the code. We refer to this special case as \emph{algebraic space-time storage code}.
\end{remark}

\section{Problem Formulation and Design Criterion for Nonparallel Channels}

It was pointed out in \cite{ICC14} that the repair/reconstruction transmissions can be modeled as a multiple access channel (MAC), where multiple users are communicating simultaneously to a joint destination. Consider now  a wireless distributed storage system with $K$ repairing\footnote{For simplicity, we will exclusively talk about repair ($K=d$), the file reconstruction process is analogous ($K=k$).} nodes, each node equipped with $n_t$ transmit antennas, and $n_r$ receive antennas at the newcomer node. Let $H_k$ be the (Rayleigh distributed) channel matrix of size $(n_r \times n_t)$ and $X_k \in \C^{n_t \times T}$ be the  code matrix transmitted by the $k$th repairing node\footnote{The authors are aware of the slight abuse of notation $k$ here, but hope that there is no danger of confusion as we will be using $K\in\{d,k\}$ to indicate the number of nodes participating in the transmission.}, where $T$ is the number of channel uses for transmission. The received signal matrix at the incoming node is given by
\begin{equation}
Y \ = \ \sum_{k=1}^K H_k X_k + W \label{eq:1},
\end{equation}
where $W$ is additive white Gaussian noise. Suppose further that after some lifting operation of the node contents, each resulting code matrix $X_k$ is taken from a rank-$s$ algebraic lattice space-time code ${\cal C}_k$ and is in the form of 
\[
X_k \ = \ \sum_{\ell=1}^s x_{k,\ell} C_{k,\ell}
\]
where $x_{k,\ell} \in \Z[\im]$; then we can rewrite \eqref{eq:1} as 
\begin{equation}
Y \ = \ \sum_{k=1}^K \sum_{\ell=1}^s H_k C_{k,\ell} x_{k,\ell} + W \label{eq:2}
\end{equation}
or equivalently in vector form
\begin{equation}
\underline{y} \ = \ H \underline{x} + \underline{w}  \label{eq:3}
\end{equation}
where $\underline{y}$ is the vectorization of matrix $Y$, $\underline{x}=[x_{1,1} \cdots x_{1,s} \cdots x_{K,s}]^\top$, and $H$ is the corresponding matrix of size $(n_r T \times K s)$ for \eqref{eq:2}. In order to have an efficient sphere-decoding algorithm for \eqref{eq:3},  we need $n_r T \geq K s$ such that after the QR-decomposition of $H=QR$, the matrix $R$ is an upper triangular matrix. This gives the criterion that the maximal number of independent QAM symbols to be sent by each repairing node is upper bounded by 
\begin{equation}
\frac{s}{T} \ \leq \ \frac{n_r}{K} \label{eq:4}
\end{equation}
in each channel use. 

The DMT-optimal MIMO multiple access channel (MAC) code \cite{kuser} has parameters $s=n_t^2 K_o$ and $T=n_t K_o$, where $K_o$ is the smallest odd integer $\geq K$. Thus by \eqref{eq:4}, such code can be efficiently sphere-decoded unless there are $n_r < K n_t$ receive antennas at replacing node. 

The design goal in this section is to provide a transmission and coding scheme for the wireless DSS which is  efficiently sphere-decodable. The performance of each scheme will be measured by the notion of diversity gain-multiplexing gain tradeoff (DMT). In particular, we say each repairing node transmits at \emph{multiplexing gain} $r$ if the actual transmission rate is $R = r \log \snr$ bits per channel uses, where $\snr$ is the signal-to-noise power ratio. Given a fixed value of multiplexing gain $r$, we say a scheme achieves diversity gain $d(r)$ if its outage probability $P_\out(r)$, which is a lower bound on the probability of decoding error, satisfies
\[
- \lim_{\snr \to \infty} \frac{\log P_\out(r)}{\log \snr} = d(r)
\]
and we will write the above as $P_{\out}(r) \doteq \snr^{-d(r)}$. 

We will compare the proposed schemes with the DMT-optimal MIMO-MAC codes presented in \cite{kuser}, which achieve the following optimal DMT
\begin{equation}
d^*_{n_t, n_r, K} (r) \ = \ \min\{ d^*_{n_t,n_r}(r), d^*_{Kn_t,n_r}(Kr)\} \label{eq:dmt}
\end{equation}
where $d^*_{m,n}(r)$ is the optimal DMT for $(n_t \times n_r)$ point-to-point MIMO channel and is given by the piecewise linear function connecting the points $(r,(n_t-r)(n_r-r))$ for $r=0,1,\cdots,\min\{n_t,n_r\}$. Notice, however, that the codes in \cite{kuser} require a high number of receiving antennas $n_r\geq Kn_t$ to achieve the optimal DMT.

\section{The Proposed Scheme : $n_t=1$, $n_r=2$, and $K$ users}
\label{protocol}

Let ${\cal K}=\{1,2,\cdots,K\}$ denote the set of $K$ users, and let ${\cal U}$ denote the set of all $2$-subsets of $\cal K$, i.e., 
\[
{\cal U}= \left\{ \{1,2\}, \{1,3,\}, \cdots, \{K-1, K\} \right\}
\]
With the above, the proposed scheme is the following. Let $U=\{u_1, u_2\}$ be a random variable uniformly distributed over $\cal U$. Given $U$, only replacing nodes $u_1$ and $u_2$ transmit during the period of $U$. Note that $\Pr\{ k \in U \} = \frac{2}{K}$ for any $k \in {\cal K}$. This means that in order to achieve an average multiplexing gain $r$, each replacing node $k$, when is chosen according to $U$, i.e. $k \in U$, should actually transmit at multiplexing gain $\frac{Kr}{2}$. Specifically, we have the following scheme:
\begin{enumerate}
\item Randomly pick $U=\{u_1, u_2\}$ from the ensemble $\cal U$.
\item The repairing nodes $u_1$ and $u_2$ transmit using the DMT-optimal MIMO-MAC code given in \cite[Eq. (20)]{kuser} for $n_t=1$, two users, and multiplexing gain $\frac{Kr}{2}$. 
\end{enumerate}
It should be noted that at each time instant, only two repairing nodes transmit to the replacing node; as $n_r = 2 n_t$ it follows from \eqref{eq:4} the scheme is efficiently sphere-decodable. An explicit example will be provided in Section \ref{first_protocol}.
\begin{figure}[t!]
\[
\includegraphics[width=2.5in]{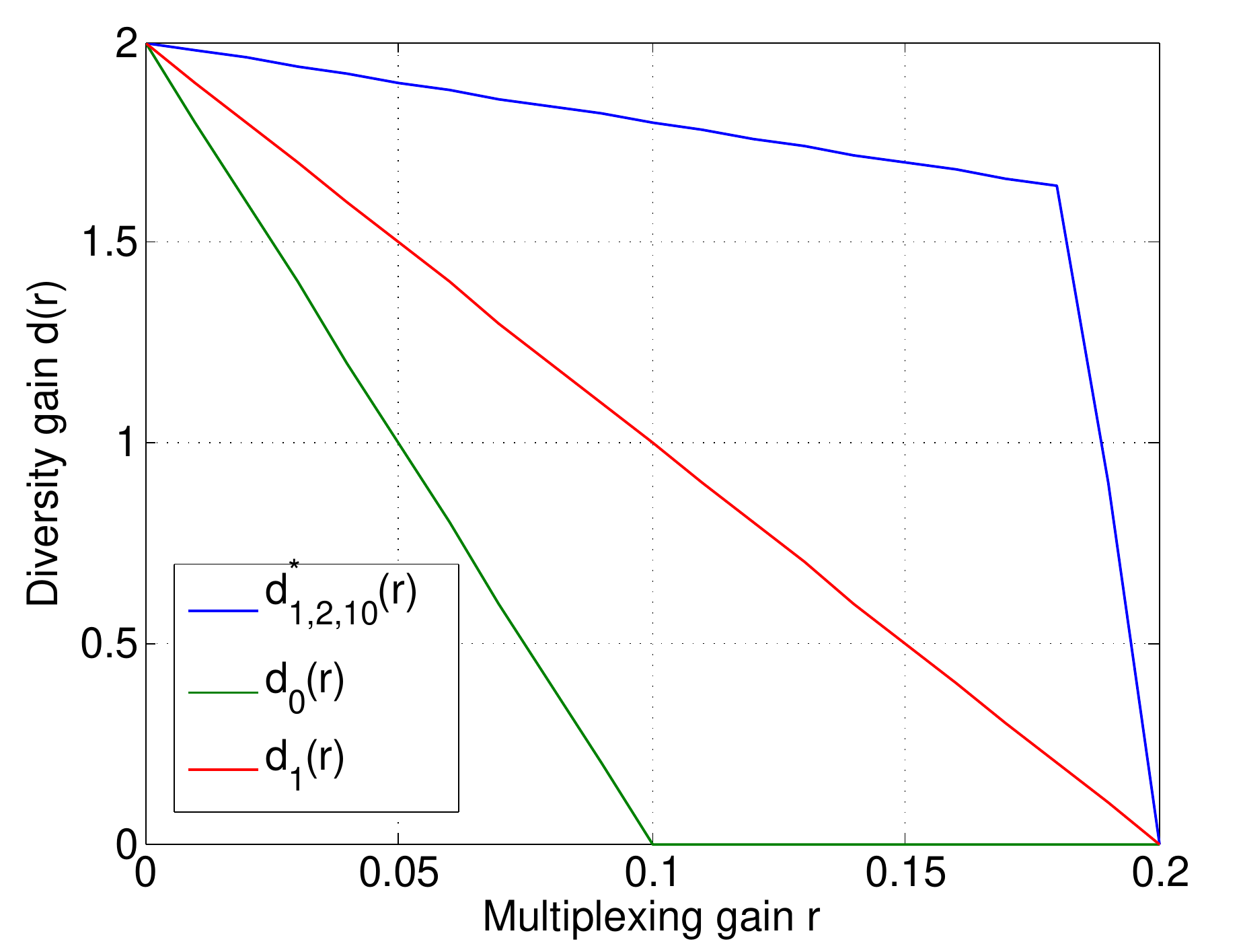}
\]
\caption{DMT Performances achieved by the optimal MIMO-MAC code, TDMA scheme, and the first proposed scheme for $K=10$ repairing nodes, $n_t=1$, and $n_r=2$.} \label{fig:1}
\end{figure}

The DMT performance achieved by this scheme is the following
\begin{equation}
d_1(r) \ = \ \min\left\{ d^*_{1,2} \left( \frac{Kr}{2} \right), d^*_{2,2} \left( Kr\right) \right\}
\end{equation}

In Fig. \ref{fig:1} we compare the DMT performance of this scheme to the DMT-optimal MIMO-MAC code, which has DMT $d^*_{1,2,10}(r)$ given in \eqref{eq:dmt}, and the DMT $d_0(r)$ of the time division multiple access (TDMA) based scheme. By this we mean that each repairing node takes turns in an orthogonal manner to transmit information to the replacing node at multiplexing gain $Kr$. It can be seen that the first proposed scheme is much better than the TDMA-based one. The gap to the optimal MAC-DMT \cite{kuser} is still big, but the optimal codes would require exponential decoding complexity (since $Kn_t>n_r$). To decode the optimal MAC space-time codes by a sphere-decoder would require $n_r\geq Kn_t$ receive antennas,  which is highly unrealistic, in particular  for  device-to-device (d2d) or peer-to-peer (p2p) networks consisting of, e.g., mobile phones and laptops having only one or two antennas.

\section{Explicit construction}
\label{first_protocol}
Let us now construct an explicit algebraic space-time storage code. We consider the case of $K$ nodes, out of which two nodes at a given time are transmitting, both having one transmit antenna, $n_t=1$.  We define the field extension  $E/F$ of degree 3, where $E=\Q(\eta = 2 \cos \left( \frac{2 \pi}{7} \right))$ and $F=\Q(\im)$, and with integral basis $\{1,\eta,\eta^2\}$ so the ring of integers $\mathcal{O}_E=\Z[\eta]$. The Galois group of $E/F$ is generated by  $\tau: \eta \mapsto \left( \eta^2 - 2\right) = 2\cos \left( \frac{4 \pi}{7} \right)$. 

Assume the nodes contain bit strings of length $m$. We use as  a lift function $L$ a restriction of a composite map $f\circ g$ consisting of the Gray map $g:\{0,1\}^m\rightarrow \Z[\im]$ which maps $m$-bit strings to the set $S$ of $2^m-$QAM symbols, and the map $f:S^3\rightarrow \OO_E^3$, where $f\left((q_1,q_2,q_3)\right)=(x,\tau(x),\tau^2(x))$ and $x=x_1+x_2\eta+x_3\eta^2$, mapping QAM-vectors to vectors in an algebraic lattice. 

The transmission matrix for each pair $\{u_1,u_2\}$ of two helper nodes is now
$$\left(\begin{array}{ccc}
x_1&\tau(x_1)&\tau^2(x_1)\\
x_2&\tau(x_2)&\tau^2(x_2)\\
\end{array}\right),
$$
where $x_i\in\OO_E$.

This scheme follows Definition \ref{ST-storage} and the protocol in Section \ref{protocol}, and hence achieves the DMT $d_1(r)$.

\section{Conclusions and future work}

In this paper, we considered space-time storage codes and a related transmission protocol which is able to maintain and repair data that lies on storage systems operating over a wireless fading channel. Here, the focus was on embedding a storage code by a suitable lifting procedure into a multiple access channel space-time code, while maintaining sphere-decodability. This take was motivated by the fact that optimal MAC space-time codes are not sphere-decodable but have exponential decoding complexity when $n_r<Kn_t$. The proposed protocol improved upon TDMA, but still falls behind the optimal MAC-DMT. Hence, future work should concentrate on improved protocols that achieve better DMT without losing tolerable decoding complexity. Also simulations should be carried out to verify the performance of the proposed protocol compared to the more straightforward combination of (optimal) regenerating codes and optimal MAC space-time codes.







\end{document}